\documentclass[preprint,aps,amsmath,amssymb,12pt]{revtex4}
\usepackage{epsfig}
\usepackage{slashed}
\usepackage{graphicx}
\usepackage{multirow,color}
\graphicspath{{PDF/}}
\begin{document}
\title{ Majorana neutrino signals at Belle-II and ILC}
\author{Chong-Xing Yue}\email{cxyue@lnnu.edu.cn}
\author{Yu-Chen Guo}\email{lgguoyuchen@126.com}
\author{Zhen-Hua Zhao}

\affiliation{
Department of Physics, Liaoning Normal University, Dalian 116029,  China
\vspace*{1.5cm}}

\begin{abstract}

For some theoretical and experimental considerations, the relatively light
Majorana neutrinos at the GeV scale have been attracting some interest.
In this article we consider a scenario with only one Majorana neutrino $N$, negligible mixing with the active neutrinos $\nu_{L}$, where the Majorana neutrino interactions could be  described in a model independent approach based on an effective theory. Under such a framework, we particularly study the feasibility of observing the $N$ with mass in the range 0$-$30 GeV via the process $e^+ e^- \to \nu N \to\gamma + \slashed E$ in the future Belle-II and ILC experiments. The results show that it is unpromising for Belle-II to observe the signal, while ILC may easily make a discovery for the Majorana neutrino.
\end{abstract}



\maketitle
\section{Introduction}

The discovery of neutrino oscillations indicates that neutrinos have non-zero masses and lepton flavors are mixed \cite{1}, which is so far the most clear experimental evidence for the existence of new physics beyond the Standard Model (SM). Various scenarios have been proposed to explain the tiny neutrino masses and the seesaw mechanism is one of the simple paradigms for generating suitable neutrino masses \cite{2}. This mechanism introduces right-handed sterile neutrinos which can have a Majorana mass as well as Yukawa couplings to the three active neutrinos. To reproduce the observed tiny neutrino masses, the Yukawa couplings must be very small.

Recently, from the theoretical and experimental points of view, there is a growing interest on the sterile neutrino with mass at the GeV scale \cite{3}. When the sterile neutrinos are lighter than the electroweak gauge boson $W$, they will behave as long-lived neutral particles with a measurable decay length, which gives us an opportunity to probe their signatures by taking advantage of the displaced vertex techniques. This fact has attracted many studies on the sterile neutrinos in the LHC \cite{4,5} and future colliders \cite{6,7}. The LHCb results about searches for massive long-lived particles decaying into $\nu jj$ \cite{8} can be used to constrain the sterile neutrino parameters \cite{9}.

Ascertaining whether neutrinos are Dirac or Majorana fermions is very important for resolving the origin of neutrino masses. As is well known, the Majorana nature of neutrinos can be revealed by the neutrinoless double beta decay ($0\nu\beta\beta$ decay) \cite{10}. Detection of Majorana neutrinos would be a signal of physics beyond the minimal seesaw mechanism leading to the well know $\nu$SM Lagrangian. In such a case the Majorana neutrino interactions could be better described in a model independent approach based on an effective theory \cite{11}, which has been tested in the LHC for its mass $M_N$ above 100 GeV \cite{12}. The LHC data at the center-of-mass energy $\sqrt{s} = 7$ TeV gives constraints on the relevant effective coupling constants. However, for $M_N < M_W$, the Majorana neutrino might be detected in the LHC and future colliders \cite{4,5,6,7,9}. Reference \cite{13} has shown that for $M_N < 30$ GeV, the dominant decay mode of the Majorana neutrino is $N\rightarrow\nu\gamma$. The values of its branching ratio are clearly larger than those for the three-fermion decay modes induced by the weak currents. This decay channel may be used to detect the Majorana neutrino via the process $e^+e^-\rightarrow \nu N\rightarrow \gamma+\slashed E$ (with $\slashed E$ being the missing energy) in the upcoming Belle-II experiment, which has been studied for the light $Z'$ gauge boson \cite{14}. The main goal of this article is to consider a scenario with only one Majorana neutrino $N$, which has a negligible mixing with the SM light neutrinos $\nu_{L}$ and interacts with $\nu_{L}$ by effective operators of higher dimension, and see whether the Majorana neutrino $N$ with a mass $M_N < 30$ GeV  can be detected via the $\gamma+\slashed E$ signal in the Belle-II experiment and future $e^+e^-$ colliders.

This article is organized as follows. In Section 2, we present the relevant effective operators given by the aforementioned effective theory and calculate the cross section of $e^+e^-\rightarrow \nu N  \rightarrow \gamma+\slashed E$ in the Belle-II experiment and future $e^+e^-$ colliders. The signal simulations and corresponding backgrounds are also discussed in this section. Our conclusions are given in Section 3.

\section{Effective couplings of Majorana neutrino and its production via $e^+e^-$ collisions}

\begin{figure}[htb]
\includegraphics [scale=0.9] {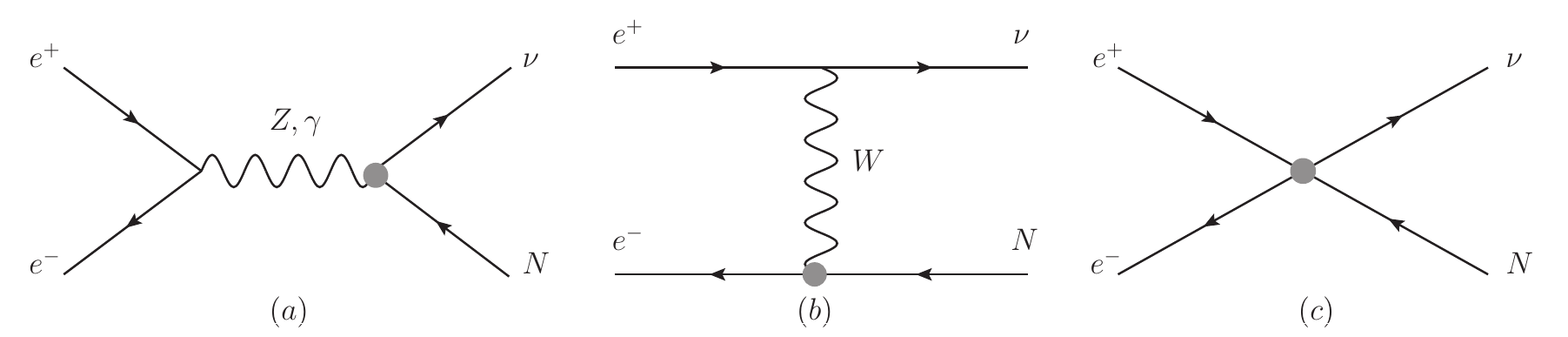}
\caption{ Feynman diagrams for the process $e^{+} e^{-}\rightarrow\nu N $. }
\label{Feynman}
\end{figure}

In this paper we consider a scenario with only one Majorana neutrino $N$ of negligible mixing with the SM neutrinos $\nu_{L}$, where the Majorana neutrino interactions could be  described in a model independent approach based on an effective theory \cite{11}. The effects of underlying new physics are parameterized by a set of effective operators that are made of the SM and Majorana neutrino fields and respect the SM $SU(2)_L \otimes U(1)_Y$ gauge symmetry \cite{15}.  The Majorana neutrino $N$ interacts with the SM neutrinos $\nu_{L}$ by effective operators of higher dimension, which is different from the traditional viewpoint that the sterile neutrinos mix with the SM neutrinos via the Yukawa couplings.

The effective operators that connect the Majorana neutrino to the SM particles can be generated at both the tree and one-loop levels \cite{11}. Among them, those which are relevant for the process $e^+e^-\rightarrow \nu N $ appear as \cite{5,13}:
\begin{eqnarray}\label{leff-tree}
 \mathcal{L}_{eff}^{tree} &=& \frac{1}{\Lambda^2}\left \{  - \alpha^{(i)}_W (\bar N_R \gamma^{\mu} l_R)\frac{v m_{W}}{\sqrt{2}}W^{+}_{\mu} + \alpha^{(i)}_{S_0}(\bar \nu_{L,i}N_R \bar e_{L,i}l_{R,i}-\bar e_{L,i}N_R \bar \nu_{L,i}l_{R,i}) \right \} + h.c.
\end{eqnarray}
and
\begin{eqnarray}\label{leff_1loop}
\mathcal{L}_{eff}^{1-loop}&=&\frac{\alpha_{L_1}^{(i)}}{\Lambda^2} \left (-i\sqrt{2} v c_W P^{(A)}_{\mu} ~\bar \nu_{L,i} \sigma^{\mu\nu} N_R~ A_{\nu}
+i \sqrt{2} v s_W P^{(Z)}_{\mu} ~\bar \nu_{L,i} \sigma^{\mu\nu} N_R~ Z_{\nu} \right )
\nonumber
\\ && -\frac{\alpha_{L_2}^{(i)}}{\Lambda^2} \left (\frac{m_Z}{\sqrt{2}}P^{(N)}_{\mu} ~\bar \nu_{L,i} N_R~ Z^{\mu}+ m_W P^{(N)}_{\mu} ~\bar l_{L,i} N_R~ W^{-\mu} \right )
\nonumber
\\ && -\frac{\alpha_{L_3}^{(i)}}{\Lambda^2}\left(i\sqrt{2} v  c_W P^{(Z)}_{\mu} ~\bar \nu_{L,i} \sigma^{\mu\nu}N_R~ Z_{\nu}
+ i\sqrt{2} v s_W P^{(A)}_{\mu} ~\bar \nu_{L,i} \sigma^{\mu\nu}N_R~ A_{\nu} \right.
\nonumber
\\ && ~~~~~~~~\left. +  ~ i \sqrt{2} v P^{(W)}_{\mu} ~\bar l_{L,i} \sigma^{\mu\nu} N_R~ W^-_{\nu} \right )
\nonumber
\\ &&  -\frac{\alpha_{L_4}^{(i)}}{\Lambda^2} \left( \frac{m_Z}{\sqrt{2}} P^{(\bar\nu)}_{\mu}~\bar \nu_{L,i} N_R~ Z^{\mu} + m_W P^{(\bar l)}_{\mu} W^{-\mu} ~\bar l_{L,i} N_R \right ) + h.c. ,
\end{eqnarray}
where $\Lambda$ is the characteristic scale for underlying physics, $\alpha$'s are the effective coupling constants, $P^{(a)}_\mu$ is the 4-momentum of $a$-particle and a sum over the family index $i$ is understood. The one-loop coupling constants are naturally suppressed by a factor $1/16\pi^2$ \cite{11,16}.

Before performing numerical calculations, we need to make clear the values of the effective coupling constants $\alpha$'s, which are restricted by the  $0\nu\beta\beta$ decay, electroweak precision data and direct collider searches. References [5, 13] have translated the existing bounds for the sterile-active neutrino mixing angles into constraints on $\alpha$'s. They have shown that the coupling constants associated to the operators that contribute to the  $0\nu\beta\beta$ decay should satisfy the relation $\alpha_{0\nu\beta\beta} \leq 3.2 \times 10^{-2} (M_{N}/100 ~{\rm GeV} )^{1/2}$, while the other ones should satisfy $\alpha \leq 0.3$ coming from the Belle experiment. In our numerical estimation, we will take their maximal values and assume the coupling constants in Eq. (2) to be $1/16\pi^2$ times the corresponding one in Eq. (1) (i.e., $\alpha^{1-loop}=\alpha^{tree}/16\pi^2$). In addition, $\Lambda$ is specified a typical value 1 TeV.

\begin{figure}[!htb]
\begin{center}
\includegraphics [scale=0.58] {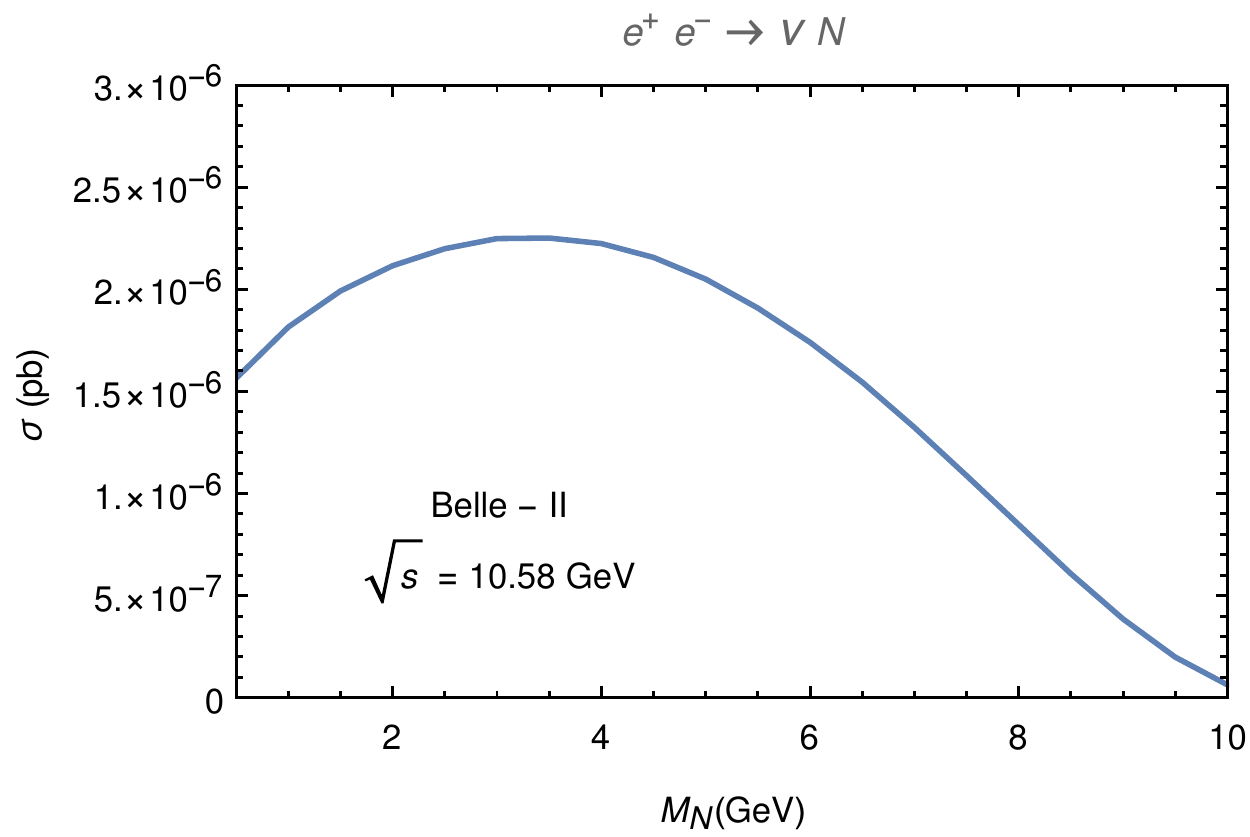}
\includegraphics [scale=0.58] {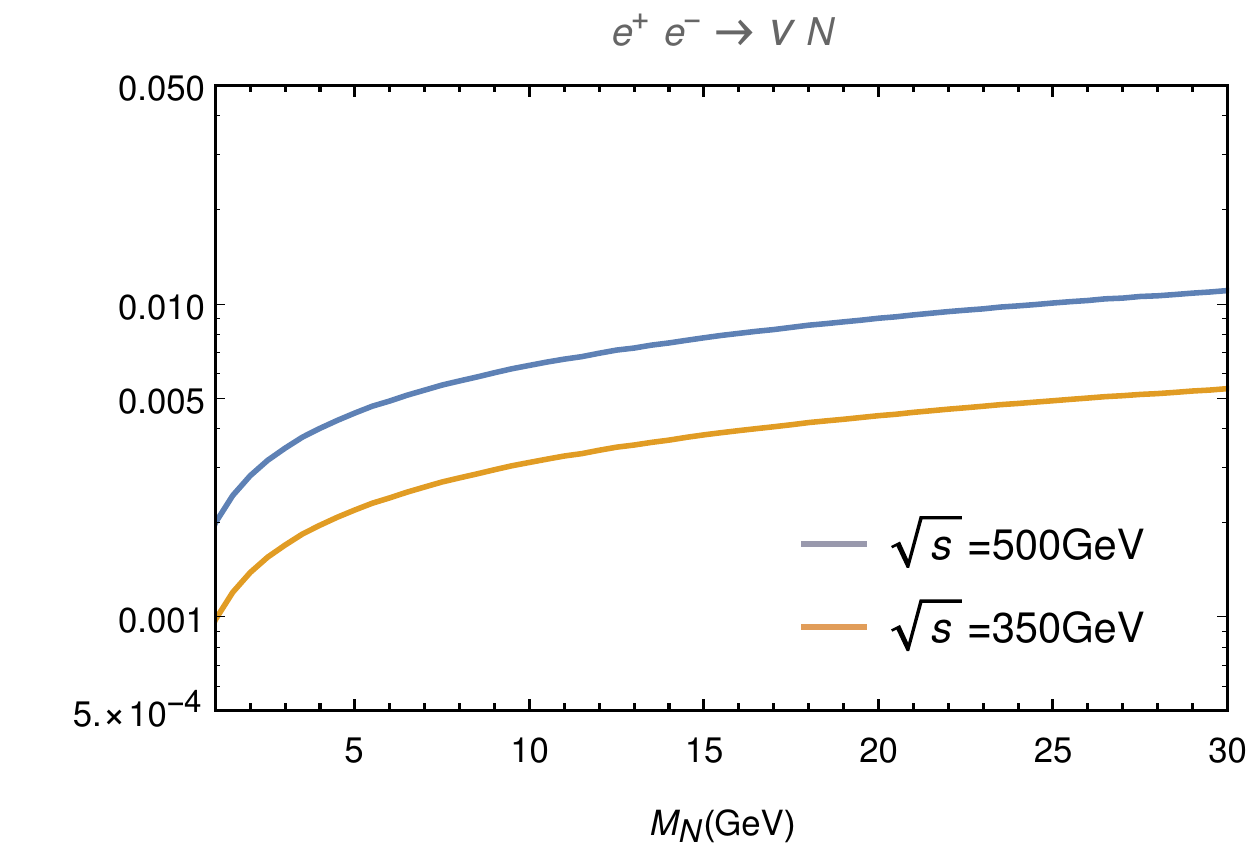}
\caption{Cross sections of $e^+e^-\rightarrow \nu N$ as a function of $M^{}_N$ at Belle-II (left) and ILC (right). } \label{ILC}
\end{center}
\label{fig2}
\end{figure}

For the scenario considered in this paper, the  mixing between the  Majorana neutrino $N$ and the SM neutrinos $\nu_{L}$ is negligibly small, and thus no operators lead to the interaction in Eq. (1) via the neutral current at tree level. The $ZN\nu$ coupling can only be generated at one-loop level in the ultraviolet underlying theory, which is suppressed by the factor $1/16\pi^2$. For $M_N < M_W$, the possible decay products of the Majorana neutrino $N$ are also three light particles, a neutrino plus a photon and light QCD-mesons, which is similar with the traditional viewpoint. Although the decay channel $N\rightarrow\nu\gamma$ is induced by the effective tensorial operators generated at loop level, Ref. \cite{13} has shown that for $M_N < 30$ GeV, it is the dominant decay mode of the Majorana neutrino. Thus, in this paper, we will focus our attention on the decay channel $N\rightarrow\nu\gamma$ and discuss the possibility of searching the  Majorana neutrino $N$ via the $\gamma+\slashed E$ signal in the Belle-II experiment and future $e^+e^-$ colliders.

We use FeynRules \cite{Feynrules} to generate the Feynman rules corresponding to the above effective operators. Then the cross section of $e^+e^-\rightarrow \nu N$ (the relevant Feynman diagrams are shown in Fig. 1) can be calculated by employing Madgraph5/aMC@NLO \cite{mg5amc}. In Fig. 2, the cross sections of this process at the future Belle-II and ILC experiments are presented as a function of $M^{}_N$. One can see that the cross section at Belle-II increases initially and decreases afterwards with a peak at $M_{N} \sim 3$ GeV, while that at ILC increases with the increase of $M_{N}$.
It should be noted that the cross section at ILC is much larger than that at Belle-II. This is because the contribution of Fig. 1(c) is greatly enhanced by a large center-of-mass energy.

\subsection{Majorana neutrino search at BELLE-II}

 We first study the feasibility of probing the Majorana neutrino at Belle-II which will operate at $\sqrt{s}= 10.58$ GeV \cite{19}. Considering the mass range possible to be explored at Belle-II, we focus on $M_{N} < 10$ GeV.

\begin{figure}[htb]
\includegraphics [scale=0.85] {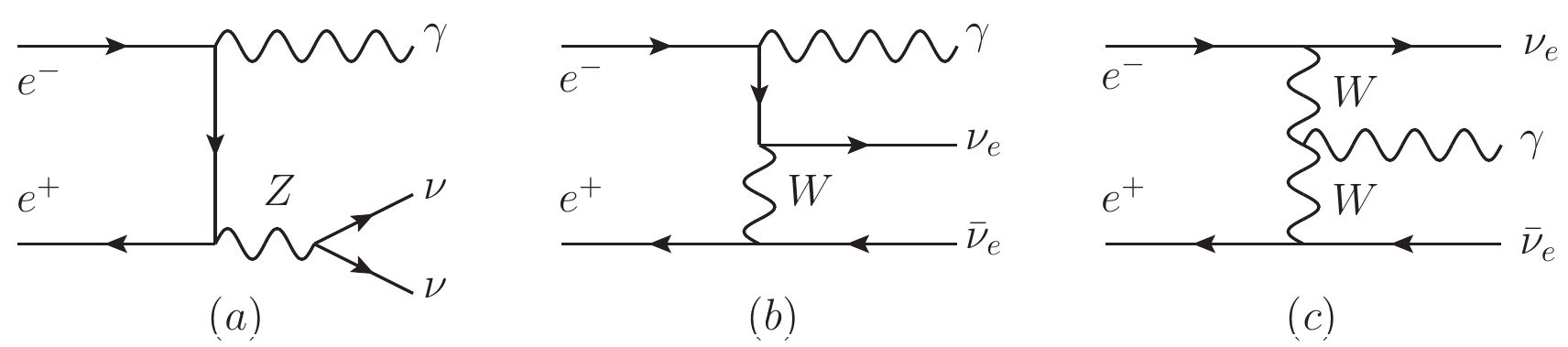}
\caption{ Feynman diagrams for the SM backgrounds of the process $e^+ e^{-} \rightarrow\gamma+ \slashed E$.}
\label{BGFeynman}
\end{figure}

The signature of the process $e^+e^-\rightarrow N\nu\rightarrow \gamma+ \slashed E$ is characterized by the presence of an isolated photon and missing transverse energy. This signal is attractive from the experimental point of view for the following two reasons. On the one hand, the experimental searches for a signal of this kind are of relatively high efficiency, since the isolated photon will be taken as the target object. On the other hand, this process will not suffer from the electromagnetic backgrounds, if the final-state particles are not missed by the detectors. However, the signal might be contaminated by the SM processes mediated by an off-shell weak boson as shown by Fig. 3. In the resulting backgrounds, the contributions of Fig. 3(a) and 3(b) dominate while the contribution of Fig. 3(c) is suppressed by the presence of an additional $W$ propagator. Note that the background processes with muon and tau neutrinos in the final state can only arise from the diagram mediated by a $Z$ boson. In comparison, all the three diagrams contribute to the background processes with electron neutrinos in the final state.

The Belle-II experiment features $\cos\theta_{\rm{min}} = 0.941$ and $\cos\theta_{\rm{max}}= 0.821$ in the center-of-mass frame, where $\theta$ is the angle between the electron beam axis and the photon momentum. Of course, such an experimental performance should be included as one of the preselection cuts. Furthermore, only the photon candidates with transverse momentum $p_T(\gamma)>$ 500 MeV will be considered in the analysis so that most of the backgrounds with soft particles
can be ruled out.

\begin{figure}[!htb]
\begin{center}
\includegraphics [scale=0.38] {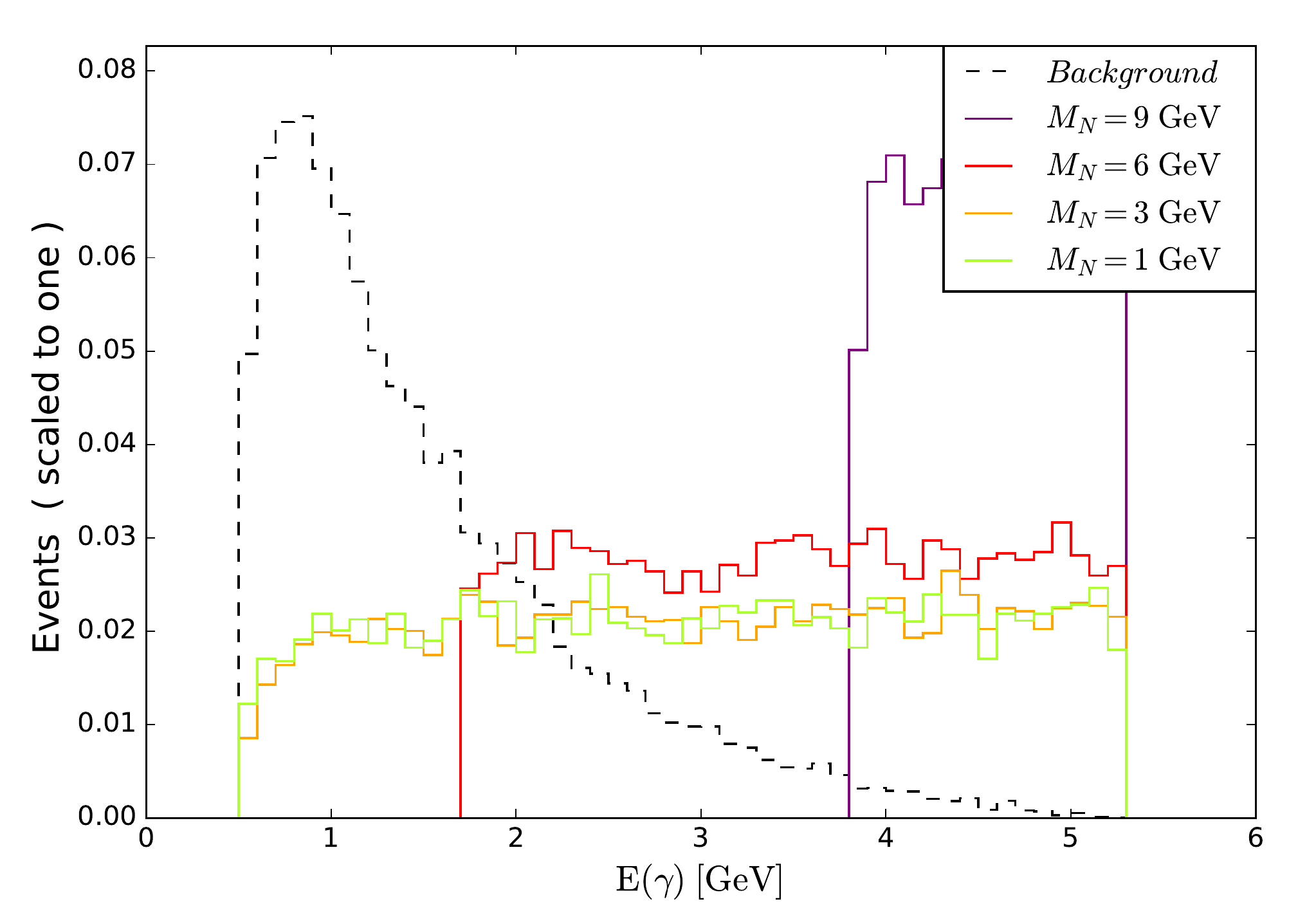}
\includegraphics [scale=0.38] {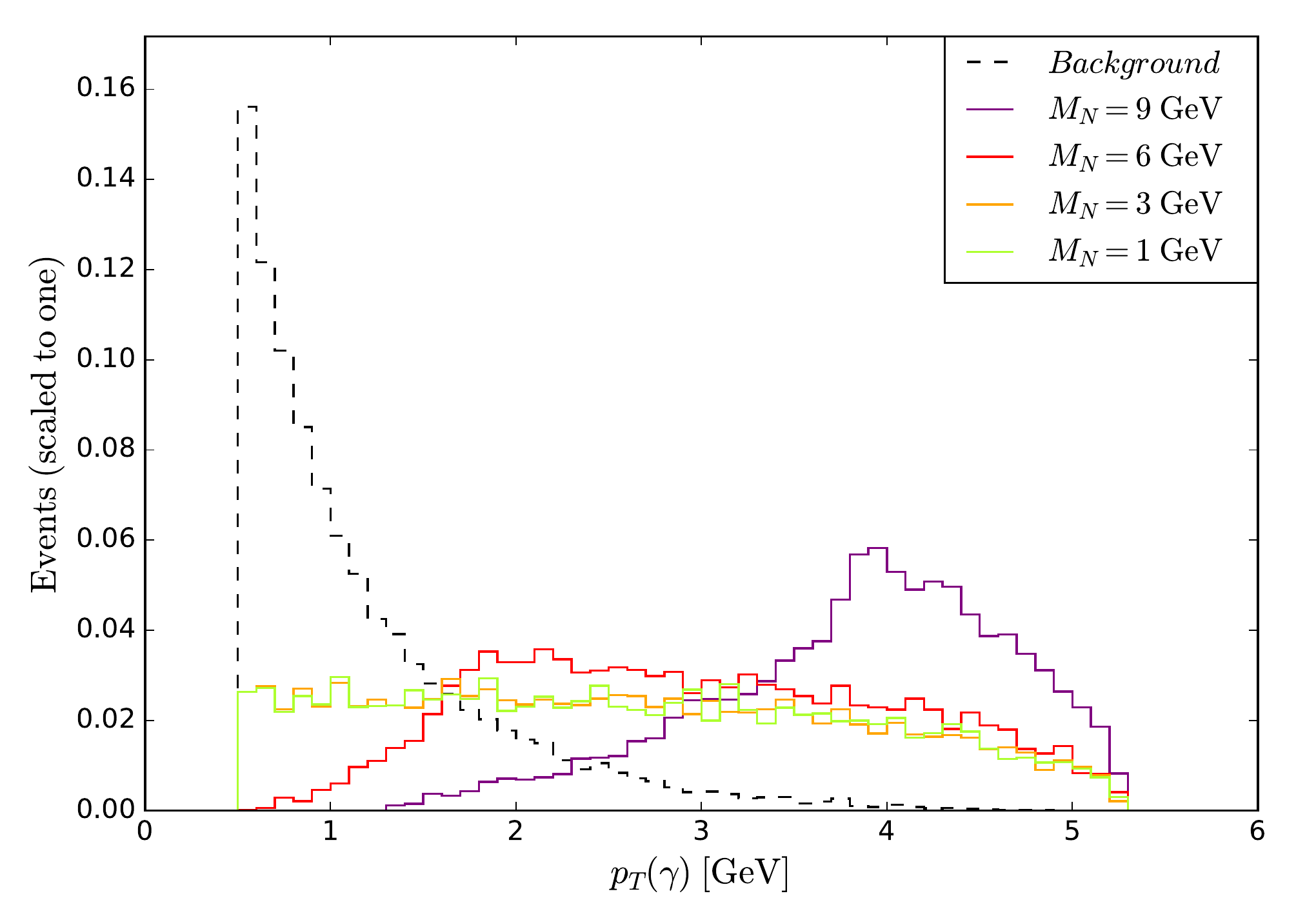}
\caption{Normalized distributions of $ E(\gamma)$ (left) and $p_{T}(\gamma)$ (right) for the signal and background photons before kinematical cuts at Belle-II.} \label{BelleDistribution}
\end{center}
\end{figure}

In Fig. \ref{BelleDistribution} we display the normalized distributions of some kinematic observables for the signal and background photons. These results are obtained from a parton level simulation under the preselection cuts by means of MadAnalysis 5 \cite{ma5}.
The left figure shows the distribution of the photon's energy $E(\gamma)$, whereas the right one for its transverse momentum $p_T(\gamma)$.

The distribution of $p_T(\gamma)$ should be same as that of the missing transverse momentum carried away by the invisible neutrinos in the final state, because the total momentum is always zero on the transverse plane. The energy and transverse momentum of signal photons are given by the mass of the Majorana neutrino and should therefore increase with $M_{N}$. Just as expected, the signal photons do have some energy and transverse momentum distributions peaking at around $M_{N}/2$ while the background photons tend to be soft. It is thus expected that the cuts on $E(\gamma)$ and $p_T(\gamma)$ in a relatively high mass region can reduce the backgrounds more efficiently. For this consideration, we introduce the improved cuts $E(\gamma)>4$ GeV and $p_T(\gamma)>4$ GeV.

\begin{table}[!htb]
\begin{center}
\caption{\small  Event numbers of the signal and backgrounds for $\mathcal{L}=50$ ab$^{-1}$ at Belle-II.}
\label{BelleII-events}
\begin{tabular}{|c|c|c|c|c|c|}\hline
   &\multicolumn{1}{|c|}{SM Background}
   &\multicolumn{1}{|c|}{$M_{N}=1$ GeV}
   &\multicolumn{1}{|c|}{$M_{N}=3$ GeV }
   &\multicolumn{1}{|c|}{$M_{N}=6$ GeV}
   &\multicolumn{1}{|c|}{$M_{N}=9$ GeV} \\
 \hline\hline
Initial                    &$4.0\times10^6$    & $90.8$          & $123$       &  $87.1$      &  $19.3$     \\ \hline
Preselection cuts          &$1.5\times10^6$    &$75.1$           & $105.42$    &  $76.48$     &  $16.95$    \\ \hline
$E(\gamma)>4{\rm GeV} $    &$2429.7$           &$21.07$          & $30.35$     &  $27.68$     &  $14.95$    \\ \hline
$p_T(\gamma)>4{\rm GeV}$   &$526.2$            &$12.7$           & $17.49$     &  $15.9$      &  $7.89$     \\ \hline
$S/\sqrt{S+B}$             &                   &$0.50$           & $0.75$      &  $0.68$      &  $0.34$     \\ \hline
\end{tabular}
\end{center}
\end{table}
We further calculate the statistical significance (SS) $S/\sqrt{S+B}$ for the signal at Belle-II, where $S$ and $B$ denote the numbers of signal and background events respectively. The integrated luminosity $\mathcal L$ is taken as 50 ab$^{-1}$ which can be achieved by the middle of the next decade. From the results in Table \ref{BelleII-events} one finds that it is unpromising to observe this signal (with a statistical significance smaller than 1).

\begin{figure}[!htb]
\begin{center}
\includegraphics [scale=0.8] {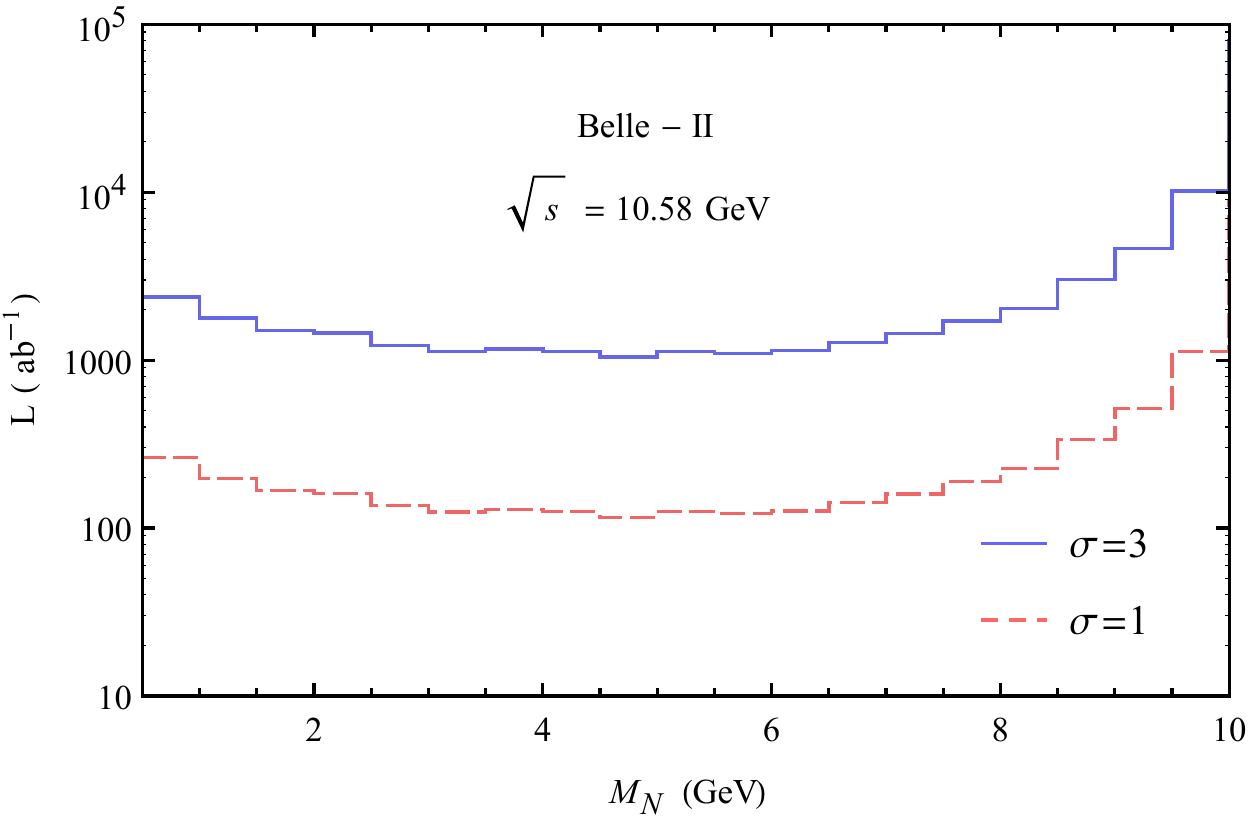}
\caption{Integrated luminosity necessary for observing the Majorana neutrino at the $1\sigma$ and $3\sigma$ levels at Belle-II.} \label{Belless}
\end{center}
\end{figure}

Finally, we give the integrated luminosity necessary for observing this signal in Fig. \ref{Belless}.
The results show that an integrated luminosity larger than $1000$ ab$^{-1}$ is essential for an observation at the 3$\sigma$ level, which far exceeds the designed luminosity of Belle-II.

\subsection{Majorana neutrino search at ILC}
We proceed to consider detecting the Majorana neutrino at ILC which will work with a center-of-mass energy of hundreds of GeV.
As a result, the mass range we will study is broaden to $M^{}_N < 30$ GeV.

To make the analysis more realistic, one needs to take into account the effects of detectors. As described in the ILC Technical Design Report \cite{ILC}, there will be two detectors and one is the Silicon Detector (SiD). With the help of PYTHIA8 \cite{py8} and Delphes3 \cite{Delphes} as well as the DSiD detector card \cite{sid}, we make a fast simulation for ILC based on the full simulation performance of the SiD.
In the simulation, the following preselection cuts are applied to the signal and background photons:
\begin{eqnarray}
  p_T(\gamma)>10 {\rm ~GeV}, \ \  |\eta(\gamma)|\!\leq\!2.5,
\end{eqnarray}
where $\eta(\gamma)$ is the pseudorapidity of photon candidates. These basic cuts are typically adopted to reproduce a general-purpose detector geometrical acceptance. After that, we further employ optimized kinematical cuts based on the kinematical differences between the signal and background photons.

\begin{figure}
\begin{center}
\includegraphics [scale=0.35] {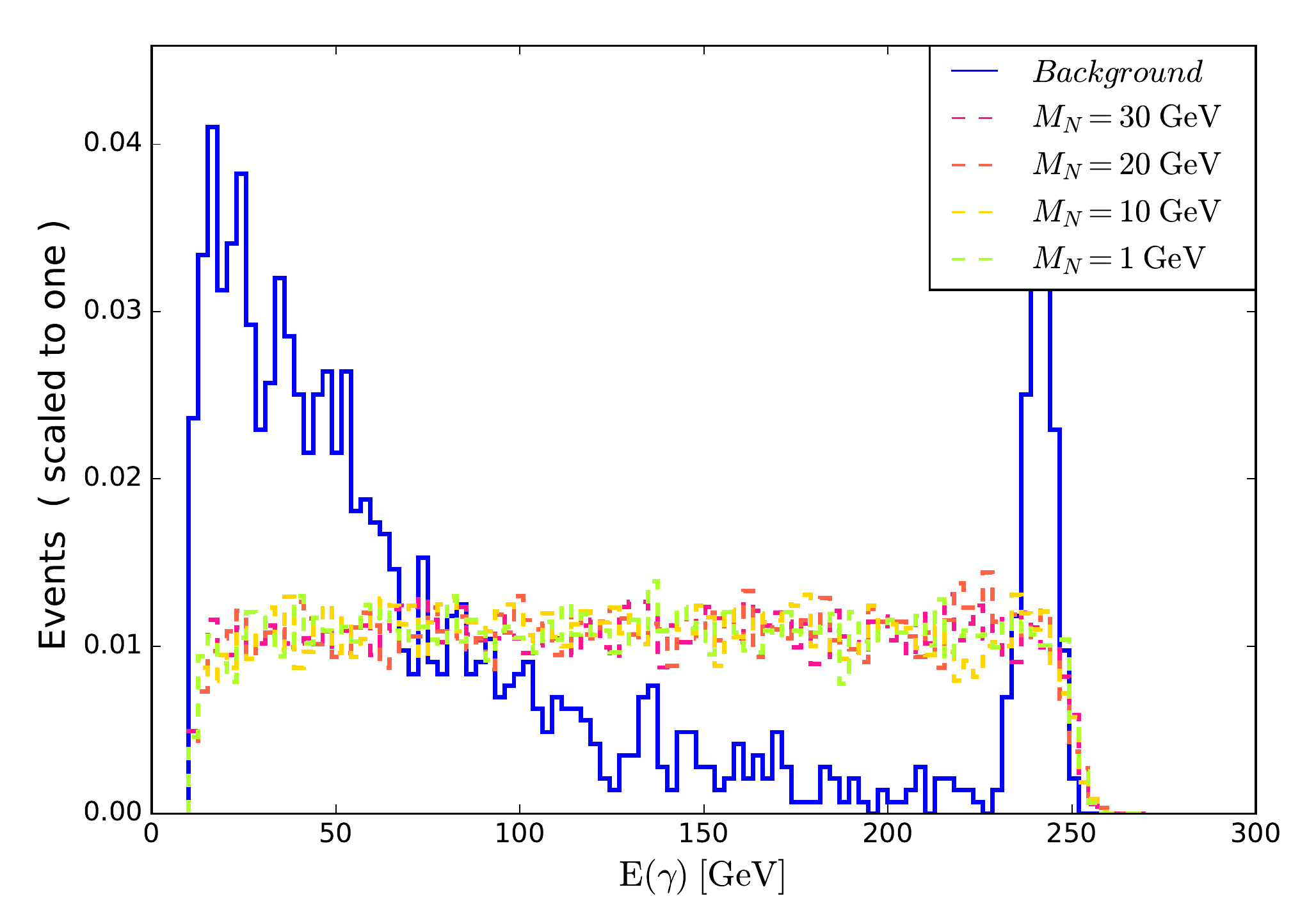}
\includegraphics [scale=0.35] {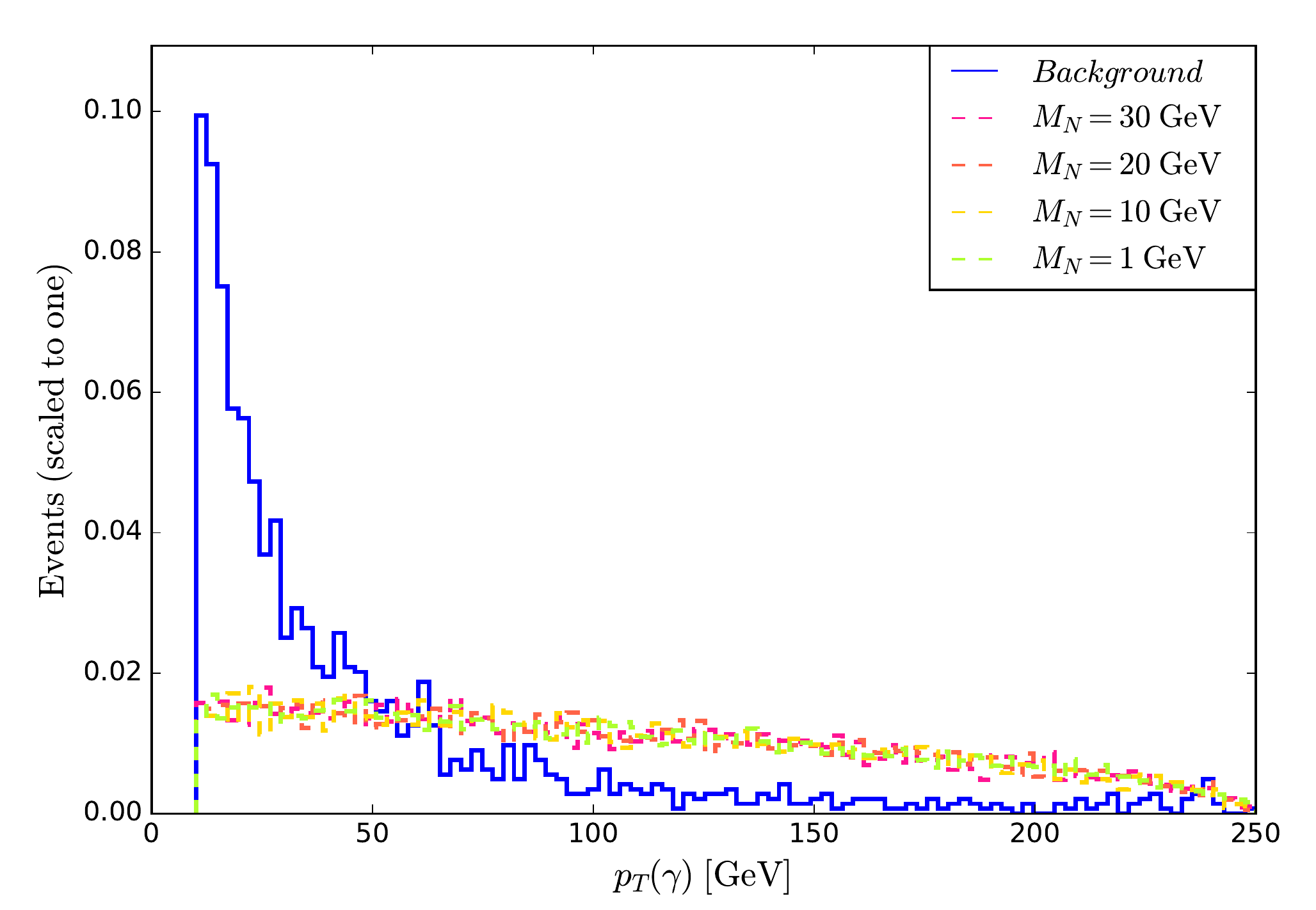}\\
\includegraphics [scale=0.35] {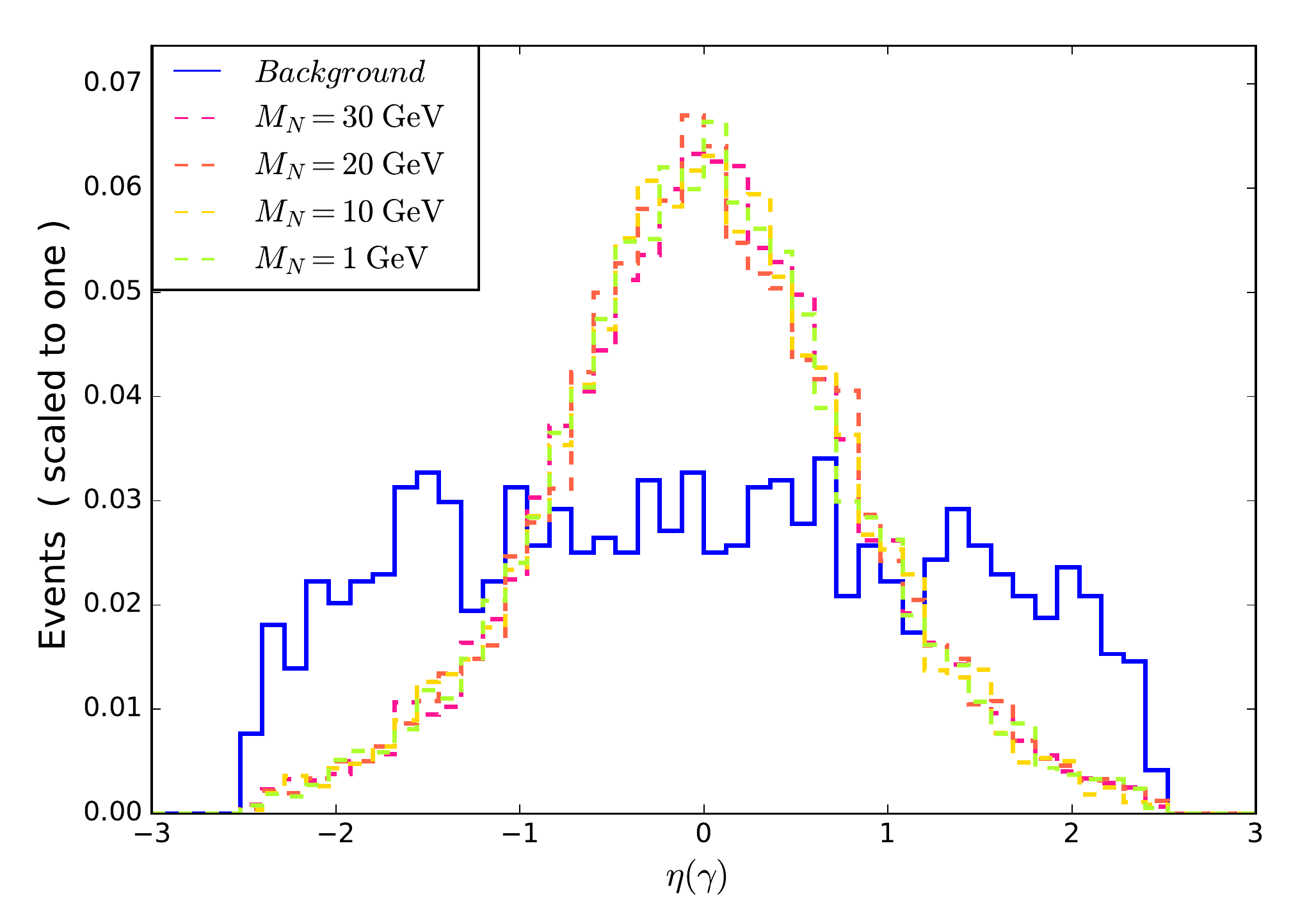}
\includegraphics [scale=0.35] {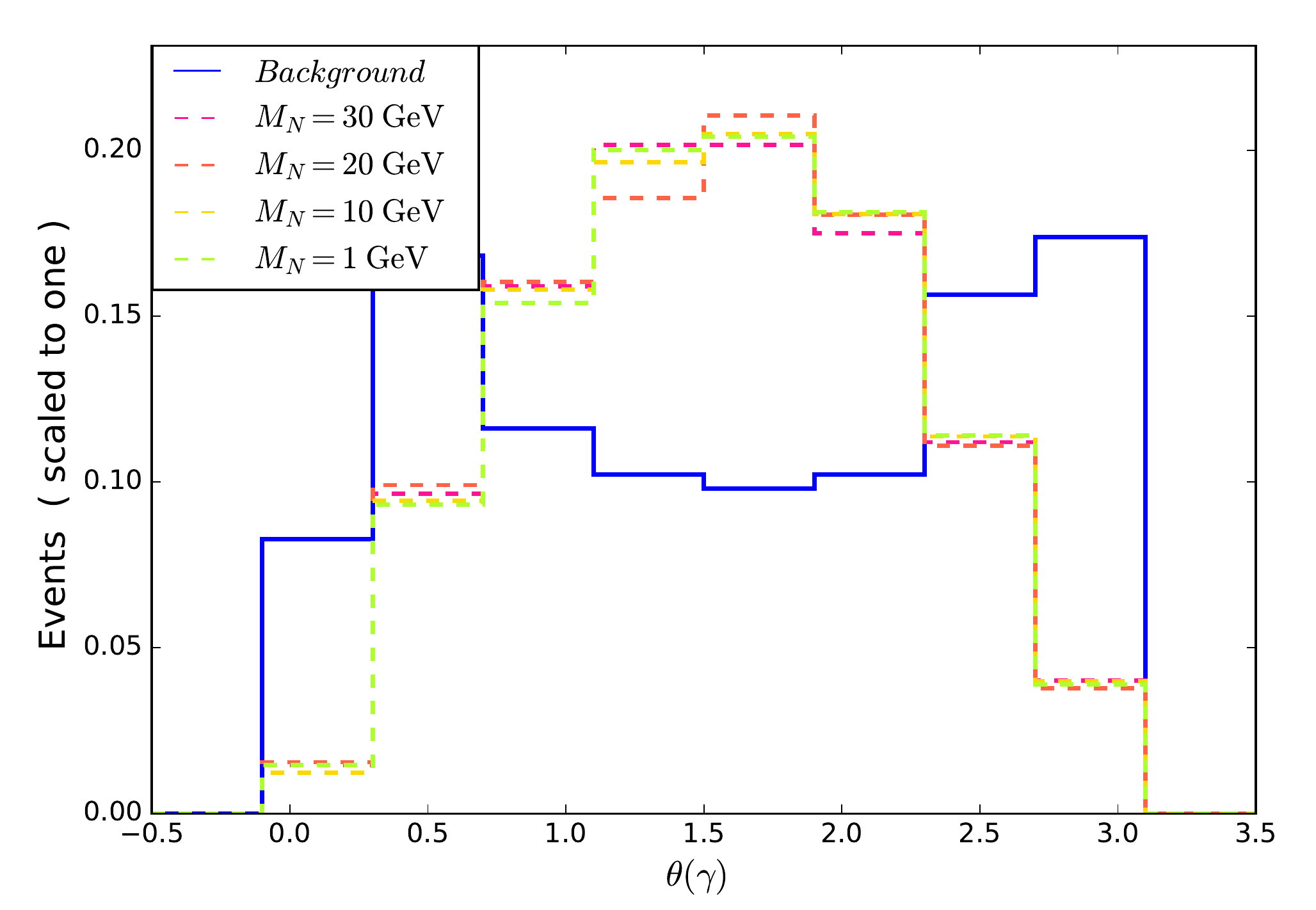}
\caption{Normalized distributions of some kinematic observables for $\sqrt{s}=$500 GeV at ILC.} \label{ILCdistribution}
\end{center}
\end{figure}

In Fig. \ref{ILCdistribution}, we display the normalized distributions of some kinematic observables for the signal and background photons at ILC with  $\sqrt{s}= 500\ \rm{GeV}$. Apparently, the signal can be well distinguished from the backgrounds by virtue of $E(\gamma)$, $p_T(\gamma)$, $\eta(\gamma)$ and $\theta(\gamma)$. Unlike at Belle-II, there is a peak near half of the center-of-mass energy in the distribution of $E(\gamma)$ for the background photons. This is attributed to the contribution of Fig. 3(a), because the $Z$ boson in such a process is on shell at ILC. The distribution of $p_T(\gamma)$ is similar to that at Belle-II in the sense that $p_T(\gamma)$ in the signal is higher than in the backgrounds. The distributions of $\eta(\gamma)$ and angle $\theta(\gamma)$ are quite convergent in the signal compared to in the backgrounds. In view of these kinematic properties, the following improved cuts will be used:
\begin{eqnarray}
  &70\ {\rm GeV} < E(\gamma) < 230\ {\rm GeV}, \ \ |\eta(\gamma)| < 1.2 ,\nonumber\\
  &0.6 < \theta(\gamma) <2.4, \ \ \ \ \ \ p_T(\gamma) > 90 {\rm ~GeV}.
\end{eqnarray}

\begin{table}[!htb]
\begin{center}
\caption{\small  Event numbers of the signal and backgrounds for $\mathcal{L}=500$ fb$^{-1}$ and $\sqrt{s}=500\ (350)$ GeV.}
\label{ILC-table}
\scalebox{0.9}{\begin{tabular}{|c|c|c|c|c|c|}\hline
   &\multicolumn{1}{|c|}{SM Background}
   &\multicolumn{1}{|c|}{$M_{N}=1$ GeV}
   &\multicolumn{1}{|c|}{$M_{N}=10$ GeV }
   &\multicolumn{1}{|c|}{$M_{N}=20$ GeV}
   &\multicolumn{1}{|c|}{$M_{N}=30$ GeV   } \\
 \hline\hline
Initial                                & $7.17(6.85)\times10^6$ & $995.0$ (487.0)& $3168$   (1551)   & $4493$   (2191)   & 5500     (2672)   \\ \hline
Preselection cuts                      & $1.03(1.04)\times10^6$ & $910.0$ (436.1)& $2912.4$ (1379.7) & $4120.3$ (1950.7) & $5040.2$ (2396.4) \\ \hline
$70{\rm GeV}<E(\gamma)<230{\rm GeV} $  & $2.67\times10^5$   & $610.1$  & $1967.2$   & $2803.7$   & $3411.1$  \\
($80{\rm GeV}<E(\gamma)<150{\rm GeV} $)& $(1.17\times10^5)$ &  (188.6) &  (582.8)   & (847.8)    &  (1045.3) \\ \hline
$|\eta(\gamma)|<1.2 $           & $8.53(6.44)\times10^4$ & $516.5$ (174.5)& $1669.3$ (534.8)  & $2362.4$ (780.1)  & $2847.8$ (964.1)  \\ \hline
$0.6<\theta(\gamma)<2.4$               & $7.17(4.80)\times10^4$ & $483.2$ (161.2)& $1572.0$ (495.7)  & $2224.9$ (723.6)  & $2683.5$ (898.6)  \\ \hline
$p_T(\gamma)>90\ (80){\rm\ GeV}$             & $3.15(2.26)\times10^4$ & $283.3$ (125.7)& $1229.5$ (391.8)  & $1766.9$ (571.0)  & $2112.0$ (717.1)  \\ \hline
$S/\sqrt{S+B}$                         &                        & $2.14$ (0.834) & $6.79$   (2.583)  &  $9.68$  (3.751)  & $11.52$  (4.696)  \\ \hline
\end{tabular}}
\end{center}
\end{table}
We subsequently calculate the statistical significance for the signal with an integrated luminosity $500$ fb$^{-1}$. The related results are summarized in Table \ref{ILC-table}, where the data out of (in) the parentheses gives the results for $\sqrt{s}=500$ GeV ($\sqrt{s}=350$ GeV). After making use of the kinematic cuts, one can gain a statistical significance larger than 5 for the signal in the case of $M_{N}>5$ GeV at $\sqrt{s}=500$ GeV.

\begin{figure}[!htb]
\begin{center}
\includegraphics [scale=0.8] {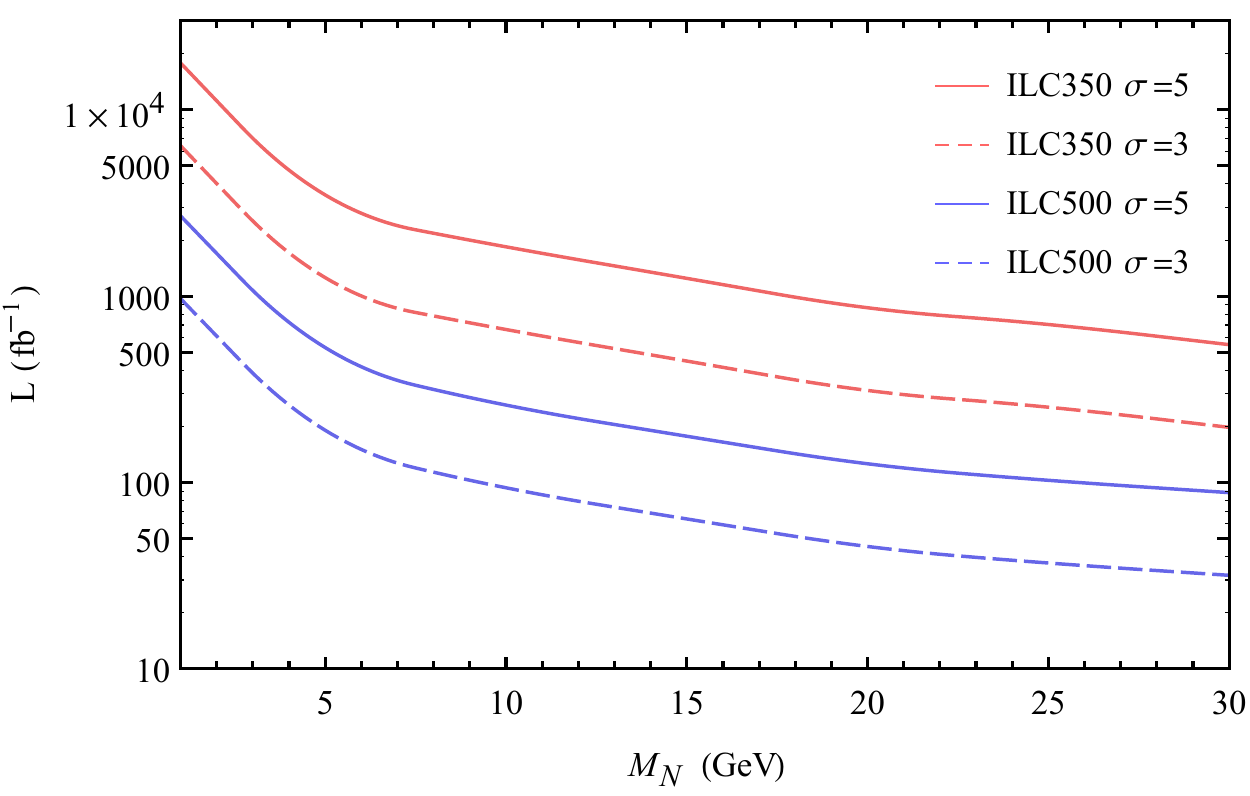}
\caption{Integrated luminosity necessary for observing the Majorana neutrino at the $3\sigma$ (dashed line) and $5\sigma$ (solid line) levels at ILC with $\sqrt{s}=350$ GeV (red) and $\sqrt{s}=500$ GeV (blue).} \label{ILC-ss}
\end{center}
\end{figure}

The integrated luminosities necessary for observing the Majorana neutrino at the $3\sigma$ and $5\sigma$ levels at ILC with $\sqrt{s}=350$ GeV and $\sqrt{s}=500$ GeV are plotted as a function of $M^{}_N$ in Fig. \ref{ILC-ss}. In light of the expected integrated luminosity 3500 fb$^{-1}$ for upgraded ILC at $\sqrt{s}= 500$ GeV, the signature of Majorana neutrinos may be easily detected in future linear colliders. But the observation capability will decrease with the decrease of $M^{}_N$.

\section{Conclusions }

The existence of Majorana neutrinos is well motivated by the famous seesaw mechanism
which allows for a natural explanation of the tiny neutrino masses. For some
theoretical and experimental considerations, the relatively light Majorana neutrinos
with masses at the GeV scale (which can therefore appear as an observable degree of freedom
at the colliders) have been attracting some interest. In this paper we consider a scenario with only one Majorana neutrino $N$ of negligible mixing with the SM neutrinos $\nu_{L}$, where the Majorana neutrino interactions could be  described in a model independent approach based on an effective theory. Under such a framework,
we particularly study the feasibility of observing the Majorana neutrino with mass
in the range 0$-$30 GeV (in which case their dominant decay mode is
$N \to \nu \gamma $) via the process $e^+ e^- \to \nu N \to \gamma + \slashed E$
in the future Belle-II and ILC experiments.

We first calculate the cross section for the production process $e^+ e^- \to \nu N$ at Belle-II and ILC.
It turns out that the cross section at ILC is much larger than that at Belle-II. Then we study the feasibility
of detecting the $\gamma + \slashed E$ signal at Belle-II and ILC by performing a signal
simulation. In order to reject the backgrounds more efficiently, some improved kinematical
cuts have been introduced based on the kinematical differences between the signal and
background photons. The results show that it is unpromising for Belle-II to observe this signal
(with a statistical significance smaller than 1), while ILC may easily make a discovery for
the Majorana neutrinos.

\section*{Acknowledgement}

\noindent
This work was supported in part by the National Natural Science Foundation of China under Grants No. 11275088 and 11605081, the Natural Science Foundation of the Liaoning Scientific Committee (No.2014020151).

\end{document}